\documentclass[11pt]{article}
\usepackage[T1]{fontenc}
\usepackage[dvips]{graphicx}
\graphicspath{{images/}}
\usepackage{verbatim}
\usepackage{amsmath,amsthm, amsfonts}
\usepackage{xspace}
\usepackage{pifont}
\usepackage{graphicx}
\usepackage{amssymb}
\usepackage{epic, eepic}
\usepackage{dsfont}
\usepackage{amssymb}
\usepackage{makeidx}
\usepackage{mathrsfs}
\usepackage{exscale}
\usepackage{color} 
\usepackage{overpic} 
\usepackage{bm}
\usepackage{bbm}
\usepackage{booktabs} 
\usepackage{color, colortbl}
\usepackage{subcaption}
\usepackage{float}
\usepackage[numbers]{natbib}
\usepackage[title]{appendix}
\usepackage{booktabs}
\usepackage{makecell}
\usepackage{enumerate}
\usepackage{multirow}
\usepackage{algorithm}       
\usepackage{algpseudocode}   

\usepackage{tikz}
\usetikzlibrary{positioning}
\usetikzlibrary {shapes,matrix}

\algtext*{EndIf}
\algtext*{EndFor}
\algtext*{EndWhile}
\algtext*{EndFunction}
\algtext*{EndProcedure}

\definecolor{crimson}{rgb}{0.86, 0.08, 0.24}

\RequirePackage[colorlinks,citecolor=blue,urlcolor=blue]{hyperref}

\definecolor{Gray}{gray}{0.9}
\usetikzlibrary{automata,positioning}

\usepackage{amsmath,afterpage}
\usepackage{epsf}
\usepackage{graphics,color}
\RequirePackage[colorlinks,citecolor=blue,urlcolor=blue]{hyperref}
\usepackage[export]{adjustbox}

\usepackage{tcolorbox}

\def\0{\mathbf{0}}

\def \< {\langle}
\def \> {\rangle}

\def\beqa{\begin{eqnarray}}
\def\eeqa{\end{eqnarray}}
\def\beqas{\begin{eqnarray*}}
\def\eeqas{\end{eqnarray*}}

\numberwithin{equation}{section}
\newcommand{\hatd}[1]{{}}

\newcommand{\bd}{\begin{displaymath}}
\newcommand{\ed}{\end{displaymath}}
\newcommand{\be}{\begin{equation}}
\newcommand{\ee}{\end{equation}}
\newcommand{\bq}{\begin{eqnarray}}
\newcommand{\eq}{\end{eqnarray}}
\newcommand{\bn}{\begin{eqnarray*}}
\newcommand{\en}{\end{eqnarray*}}

\usepackage[normalem]{ulem}

\usepackage{mathtools}

\mathtoolsset{showonlyrefs}

\usepackage{authblk}

\title{FX Market Making with Internal Liquidity}
\author[1]{Alexander Barzykin }
\author[1, 2]{Robert Boyce}
\author[2]{Eyal Neuman }
\affil[1]{HSBC}
\affil[2]{Department of Mathematics, Imperial College London }

\begin{document}

\vspace{-0.5cm}
\maketitle

\begin{abstract}
\noindent
As the FX markets continue to evolve, many institutions have started offering passive access to their internal liquidity pools. 
Market makers act as principal and have the opportunity to fill those orders as part of their risk management, or they may choose to adjust pricing to their external OTC franchise to facilitate the matching flow. 
It is, a priori, unclear how the strategies managing internal liquidity should depend on market condions, the market maker's risk appetite, and the placement algorithms deployed by participating clients. 
The market maker's actions in the presence of passive orders are relevant not only for their own objectives, but also for those liquidity providers who have certain expectations of the execution speed. 
In this work, we investigate the optimal multi-objective strategy of a market maker with an option to take liquidity on an internal exchange, and draw important qualitative insights for real-world trading.
\end{abstract} 



\section{Introduction} \label{sec:intro}

Internalisation has long been a key component of efficient algorithmic execution in foreign exchange (FX) markets, primarily due to its reduced visibility and consequently minimal market impact. Traditional internalisation typically involves client-to-client matching, where a liquidity provider acts as a neutral intermediary. This mechanism, classified as Internalisation Type A by the Foreign Exchange Professionals Association (FXPA) \cite{FXPA}, offers certain advantages but also inherent limitations, as matching opportunities tend to be scarce.

Since FX trading remains largely over-the-counter (OTC), interaction with OTC liquidity is expected to be the primary driver of internalisation. The FXPA defines Internalisation Type B as the offsetting of commercial flow by a liquidity provider. This naturally involves client algorithms trading against a market maker’s pricing stream, benefitting from ultra-low latency and potentially tighter spreads due to inventory skew.

Recently, several institutions have begun offering passive access to internal liquidity, either through conventional limit or pegged orders. Dynamic orders are typically pegged to an internally maintained fair reference price, allowing client algorithms to communicate in the “high-frequency language” of the market maker, without requiring high-frequency order management. Market makers may fill these orders to meet their risk management objectives; importantly, they can also adjust pricing in their OTC franchise to facilitate such fills. In this way, client algorithms can interact with deep OTC liquidity through the internalisation mechanism.

One might tacitly assume that market makers directly transfer pegged order quantities onto their pricing ladder at the same price and then immediately fill them upon receiving an opposing trade. However, such a naïve approach would clearly be detrimental to the market maker’s risk management and would reduce potential P\&L. In practice, market makers must instead solve an optimal market-making problem in the presence of an additional liquidity source, be it a single limit order or, more generally, a limit order book on an internal exchange.

Understanding the underlying mathematical formulation of this problem can improve the transparency of internalisation, in line with recent FXPA guidance. Orders on the internal exchange cannot be taken for granted: they may be cancelled, be of finite but unknown size, or follow an unpredictable strategy and thus may not always be available. Moreover, the market-making desk may prefer to fill client orders on the internal exchange, as doing so supports the firm’s client algo desk, which manages those orders, and ultimately provides better service to clients. The market-making desk therefore faces a complex and multifaceted trade-off.

The mathematical finance literature on market making originates from the model of \citet{avellaneda2008high}, later solved in closed form by \citet{gueant2013solution}. Subsequent extensions include the linear–quadratic framework of \citet{cartea2015algorithmic}, influential order effects in \citet{cartea2014buy}, continuous hedging in \citet{barzykin2023algorithmic}, and competition among market makers in \citet{boyce2025competition}. Interactions between market makers and clients have also been explored, notably through the game-theoretic approach of \citet{cartea2025informed}, while internalisation has been examined in various contexts, including the FX market in \citet{butz2019internalisation}.
Since \citet{avellaneda2008high}, financial markets and the role of market makers have evolved substantially. A key recent innovation is the emergence of internal exchanges, enabling clients to provide liquidity directly to market makers. This remains largely unexplored, with the exception of \citet{morimoto2024internal}, who study optimal execution under unlimited internal liquidity with price impact. Relatedly, passive impact has attracted renewed attention, most notably in \citet{chahdi2024passive}.

In this work, we present the first quantitative investigation of internal exchange management from the market maker’s perspective. We develop a model in which the market maker continuously streams a price ladder of multiple sizes to external, liquidity-taking clients while optimally timing trades with liquidity-providing clients on an internal exchange. The market maker aims to maximise P\&L while remaining averse to large inventory positions and unfilled internal orders. Internal exchange orders are transient, that is they may be cancelled, executed, or reappear later, capturing realistic client execution patterns such as iceberg, TWAP, or full-amount strategies. From a mathematical perspective, our formulation of the market maker's problem leads to a stochastic control problem over external prices, consistent with classical market-making frameworks \citep{avellaneda2008high,bergault2021size}, combined with repeated optimal stopping decisions for the timing of internal trades, as in the optimal execution framework of \citep{cartea2015limit,morimoto2024internal}. The resulting Hamilton–Jacobi–Bellman quasi-variational inequality (HJBQVI) is then solved numerically.

We derive practical insights with direct relevance to real-world trading. In particular, we demonstrate that there exists an execution threshold, i.e. the inventory level beyond which the market maker will instantaneously fill the limit order. Otherwise, the market maker will adjust pricing to external OTC clients accelerating the move towards the execution threshold, thus facilitating the fill of the limit order. The degree of price skew and the execution threshold level depend on the market maker's risk aversion, limit order depth and expected placement strategy, as well as on flow facilitation initiatives. Importantly, the optimal strategy significantly outperforms a naïve benchmark strategy that directly incorporates internal exchange orders into the OTC pricing ladder.

\section{The model} \label{sec:model}

Let $T>0$ denote the trading period. We fix a filtered probability space $(\Omega, \mathcal{F}, \mathbb{F}=\{\mathcal{F}_{t}\}_{t\in[0, T]}, \mathbb{P})$ satisfying the usual hypothesis. Let $(S_{t})_{t\geq 0}$ be the price process of a risky asset such that $S_t=S_{0}+\sigma W_{t}$, where $W$ is a standard Brownian motion and $S_0, \, \sigma$ are positive constants.  We consider a market maker who continuously provides liquidity on both sides of the order book, quoting bid and ask prices $(S^{b, z}, S^{a, z})$ that are streamed and adjusted according to the clients' order sizes  $\mathcal{Z}=\{ z_k,\, 1 \leq k \leq K \}$: 
  \begin{equation} \label{def:S_a_b}
	S^{b, z}_{t} = S_{t} - \delta^{b, z}_{t}
	\qquad \text{and} \qquad
	S^{a, z}_{t} = S_{t} + \delta^{a, z}_{t}, \quad 0\leq  t  \leq T, \quad z \in \mathcal{Z}. 
\end{equation}
Note that the half-spreads $(\delta^{b, z},  \delta^{a, z})$ are controlled by the market maker and chosen from the set of admissible half-spreads
\begin{equation} \label{def:delta_admissiblity} 
    \mathcal{D} = \left\{ \delta: \delta \text{ progressively measurable s.t. } \mathbb{E}\left[\int_{0}^{T}\delta^{2}_{t}dt\right]<\infty \right\}.
\end{equation}
Market buy and sell orders of OTC clients are modelled by independent counting processes $N^{a, z}=(N_t^{a, z})_{t\geq 0}$ and $N^{b, z}=(N_t^{b, z})_{t\geq 0}$  for any order size $z\in\mathcal{Z}$, with intensities 
\begin{equation} \label{def:lambda_a_b}
	\Lambda^{b/a, z}(\delta^{b/a, z}_{t}) = \lambda^{b/a,  z}\exp\Big(-\kappa^{z}\delta^{b/a, z}_{t}\Big)
	\qquad 0\leq  t  \leq T.
\end{equation}
Here $\lambda^{b/a, z}$ and $\kappa^{z}$ are positive constants. 

We assume that, in the internal exchange, the client has placed a limit order on one side of the book. Without loss of generality, we take this to be an ask limit order with an initial size of $\overline{\ell}$; that is, the client intends to sell, and the dealer would buy if a trade occurs.  

The order size at any time $t$, denoted by $\{L_{t}\}_{0,T}$ which is a càdlàg process that determines the liquidity in the internal exchange and it is given by, 
\begin{equation} \label{def:L}
    L_{t} =\overline{\ell}-\overline{\ell}\int_{0}^{t}\mathbbm{1}_{\{L_{s}>0\}}dC_{s}+\overline{\ell}\int_{0}^{t}\mathbbm{1}_{\{L_{s}\leq0\}}dA_{s}+\overline{\ell} R_{t} - M_{t}, \quad 0\leq  t  \leq T,
\end{equation}
Note that $\overline{\ell} > 0$ corresponds to an order being present at time $t = 0$, whereas $\overline{\ell} \leq 0$ indicates the absence of an order. Moreover, larger magnitudes of $\overline{\ell}$ decrease the likelihood of the order appearing, since liquidity can be consumed by the market maker only when $L_t > 0$ in \eqref{def:L}. The processes $(A, C, M, R)$ are defined and characterized below.
\begin{itemize}
   \item $(C_t)_{t\geq 0}$ represents the cancellation of the order by the client and is modelled as an independent Poisson process with intensity $\nu$ and unit jump size. An order can be cancelled by the client only if it is present, so jumps of $C$ affect $L$ only when liquidity is available at that time.

    \item $(A_t)_{t\geq 0}$ represents the arrival of new orders and may also be interpreted as the non-instantaneous replenishment of a previously filled limit order. It is modelled as a Poisson process with intensity $\mu$ and unit jump size. While, in practice, a new order could arrive while another is still active, such events are sufficiently rare that we restrict arrivals to occur only when no order is currently present. This assumption is consistent with the behaviour of a TWAP placement algorithm.

\item $
(M_t)_{t \geq 0}
$
represents the cumulative market orders submitted by the dealer. We assume that these orders occur at jump times, which are $\mathbb{F}$-stopping times $\{\tau_n\}_{n \geq 1}$ controlled by the market maker, and that each transaction is of unit size. Market orders can only occur at times when $L_t > 0$, and are zero otherwise. Hence, they are chosen from the class  
\begin{equation} \label{def:time_admissibilty}
\mathbb{T} = \left\{ \tau : \tau \text{ is an } \mathbb{F}\text{-stopping time and } L_{\tau-} \geq 1 \right\}.
\end{equation}
We then define  $
M_t = \sum_{n \ge 1} \mathbbm{1}_{\{\tau_n \leq t\}}.  $
\item\((R_t)_{t \geq 0}\) represents the replenishment process of an order immediately after the dealer consumes the last unit of liquidity. We assume that $
R_{t} = \sum_{n=1}^{\infty} \chi_{n} \mathbbm{1}_{\{\tau_{n} \leq t\}} \mathbbm{1}_{\{L_{\tau_n}=0\}},$
where \((\chi_{n})_{n \geq 1}\) is a sequence of i.i.d.\ Bernoulli random variables with parameter \(p \in [0, 1]\). Replenishment typically occurs when an internal exchange order is part of a larger iceberg order.

\end{itemize}
The pricing offset of the internal exchange order relative to the mid-price is denoted by the parameter \(\rho\), which can be positive or negative. The price at which the market maker can trade is given by, 
\begin{equation}
    P_{t} =
    \begin{dcases}
        S_{t}+\rho & \text{ if } L_{t}>0 \\
        \infty & \text{ if } L_{t} \leq 0. 
    \end{dcases}
\end{equation}
The dealer receives an infinitely unfavorable price when trading is impossible. In practice, internal exchange orders may yield a small fee for the client. From the modeling perspective this can be incorporated by letting \(\rho = \tilde{\rho} - \xi\), where \(\tilde{\rho}\) is the price offset relative to the mid chosen by the client, and \(\xi > 0\) is a constant representing the fee per unit. Throughout the paper, we work directly with \(\rho\).

The dealer's position and cash processes are given by, 
\begin{equation}
 Q_{t} = \sum_{z\in\mathcal{Z}}z\left( N^{b, z}_{t} - N^{a, z}_{t}\right) + M_{t},
 \end{equation}
 \begin{equation}
 X_{t} = \sum_{z\in\mathcal{Z}}z\left(\int_{0}^{t}S^{a, z}_{s}dN^{a, z}_{s} - \int_{0}^{t}S^{b, z}_{s}dN^{b, z}_{s}\right) - \int_{0}^{t}P_{s}dM_{s},
\end{equation}
respectively. The value function of the maker maker is, 
\begin{equation} \label{def:value}
\begin{split}
     v(t, s, q, x, l) 
     =   \sup_{ \delta^{b},\delta^{a},(\tau_{n})_{n\geq 1}}\mathbb{E}\Bigg[X_{T} + Q_{T}S_{T} - \alpha Q^{2}_{T} - \phi\int_{t}^{T}Q^{2}_{s}ds - \psi\int_{t}^{T} (L_{s})^+ds \bigg\vert\,\mathcal{F}_{t}\Bigg],
\end{split}
\end{equation}
were the supremum is taken over $ \delta^{b},\delta^{a} \in \mathcal{D}$ and $\tau_{n}\in \mathbb{T}$ (see \eqref{def:delta_admissiblity} and \eqref{def:time_admissibilty}). The constants $\alpha, \phi, \psi$ are nonnegative and $(\cdot)^+$ is the positive part function. The first two terms on the right-hand side of \eqref{def:value} represent the terminal value of the market maker's portfolio; that is, the cash position plus the risky asset position valued at mid. The third and fourth terms implement penalties on the terminal and running positions respectively. The fifth term implements a running penalty on unfilled internal exchange orders. 

Using the mathematical argument in \cite[Chapter 11,Theorem 11.1]{oksendal2019jump}, the dynamic programming principle yields that the value function $v$ satisfies the following Hamilton-Jacobi-Bellman quasi-variational inequality (HJBQVI), 

\begin{equation} \label{eq:HJBQVI}
  \begin{aligned} &0=\max \Bigg\{ \frac{\partial v}{\partial t}(t, s, q, x, l) + \frac{\sigma^{2}}{2}\frac{\partial^{2} v}{\partial s^{2}}(t, s, q, x, l) - \phi q^{2} - \psi \cdot (l)^+ \\ &\quad +\sum_{z\in\mathcal{Z}, i\in\{b, a\}}\left(\underset{\delta^{i, z}}{\sup}\left(\lambda^{i, z}e^{-\kappa^{z}\delta^{i, z}}\left(v(t, s, q \pm z, x-zs+z\delta^{i, z}, l) - v(t, s, q, x, l)\right)\right)\right) \\ &\quad+\nu\left(v(t, s, q, x, l-\overline{\ell}) - v(t, s, q, x, l)\right)\mathbbm{1}_{\{l>0\}} +\mu\left(v(t, s, q, x, l+\overline{\ell}) - v(t, s, q, x, l)\right)\mathbbm{1}_{\{l\leq0\}} \mathbf{,} \\ &\qquad \mathbbm{1}_{\{l>1\}}\left(v(t, s, q+1, x-s-\rho, l-1)\right) \\ &\quad+\mathbbm{1}_{\{l\leq1\}}\big(p\,v\left(t, s, q+1, x - s - (\rho\mathbbm{1}_{\{l>0\}} + \infty\mathbbm{1}_{\{l\leq0\}}), l-1+\overline{\ell}\right) \\ &\qquad\quad+(1-p)\,v\left(t, s, q+1, x - s - (\rho\mathbbm{1}_{\{l>0\}} + \infty\mathbbm{1}_{\{l\leq0\}}), l-1\right)\big) - v(t, s, q, x, l) \Bigg\},\end{aligned}
	\end{equation}
with terminal condition
\begin{equation} \label{eq:terminal_v}
	    v(T, s, q, x, l) = x + qs - \alpha q^{2}.
\end{equation}
 The terms on the first argument of the maximum in \eqref{def:value} arise from It\^{o}'s formula for jump diffusions. The second part of the maximum relates to times where an internal exchange order is traded. In particular, when there is no internal exchange standing liquidity, this term evaluates to $-\infty$ and thus the first part of the maximum is larger. The $\pm$ sign in \eqref{eq:HJBQVI} relates to terms indexed by $b$ ($a$) having a plus (minus) sign, respectively.  
By using the ansatz 
\begin{equation} \label{eq:ansatz}
	v(t, s, q, x, l) = x + qs + h(t, q, l),
\end{equation}
we can reduce the dimension of \eqref{eq:HJBQVI}. Solving the Hamiltonians in feedback form then yields
 	\begin{equation} \label{eq:HJBQVI_simplified}
	\begin{split}
	    0 = \max\Bigg\{
	    &\frac{\partial h}{\partial t}(t, q, l) - \phi q^{2} - \psi\cdot (l)^+ \\
	    &+\sum_{z\in\mathcal{Z},i\in\{b, a\}}\left(\frac{z\lambda^{i, z}e^{-1}}{\kappa^{z}}\exp\left(\frac{\kappa^{z}}{z}
	    \left( h(t, q\pm z, l) - h(t, q, l)\right)\right)\right) \\
	    &+\nu\left(h(t, q, l-\overline{\ell}) - h(t, q, l)\right)\mathbbm{1}_{\{l>0\}} + \mu\left(h(t, q, l+\overline{\ell}) - h(t, q, l)\right)\mathbbm{1}_{\{l\leq0\}}, \\
	    &\left(p\,h(t, q+1, l-1+\overline{\ell}) + (1-p)\,h(t, q+1, l-1)\right)\mathbbm{1}_{\{l\leq1\}} \\
	    &+ h(t, q+1, l-1)\mathbbm{1}_{\{l>1\}} - h(t, q, l) - \left(\rho\mathbbm{1}_{\{l>0\}} + \infty\mathbbm{1}_{\{l\leq0\}}\right)
	    \Bigg\}
	\end{split}
	\end{equation}
with terminal condition $h(T, q, l) = -\alpha q^{2}$.
The optimal depths are given by, 
	\begin{equation} \label{eq:deltas}
	\begin{aligned}
   		&\delta^{i, z}(t, q, l) = \frac{1}{\kappa^{z}} + \frac{1}{z}\left(h(t, q, l) -  h(t, q \pm z, l)\right),
        \qquad \text{ for } i\in\{b, a\},
	\end{aligned}
	\end{equation}
and the optimal execution times $(\tau_n)_{n\geq 0}$ are the times such that the maximum in \eqref{eq:HJBQVI_simplified} evaluates as the second term. We refer to the subspace where this holds as the execution region.

\section{Optimal strategy} \label{sec:strategy}

In this section, we investigate the behaviour of the optimal strategy obtained by numerically solving \eqref{eq:HJBQVI_simplified} using a backward Euler scheme. 
An anonymised subsample of HSBC GBPUSD trade data was used to calibrate the intensities $\lambda^{a, z}$ and $\lambda^{b, z}$, as well as the sensitivity of fill probabilities to quotes, $\kappa^{z}$, for $z \in \mathcal{Z} = \{1, 5, 10\}$. 
Specifically, we set $\kappa^{1} = 1.5$, $\kappa^{5} = 1.0$, $\kappa^{10} = 0.5$, $\lambda^{b, 1} = \lambda^{a, 1} = 0.2$, $\lambda^{b, 5} = \lambda^{a, 5} = 0.005$, and $\lambda^{b, 10} = \lambda^{a, 10} = 0.001$. 
Each $\kappa$ is measured in bps\textsuperscript{-1}, and each $\lambda^{a, z}$, $\lambda^{b, z}$ in seconds\textsuperscript{-1}. 
The inventory penalties are given by $\phi = \alpha = 0.001$, and the penalty for unfilled internal exchange orders is $\psi = 0.01$. 
The time horizon is $T = 300$ seconds, which represents a reasonable high-frequency risk management period. 
We consider a range of client prices $\tilde{\rho}$ and three parameter configurations for \eqref{def:L} corresponding to different client algos. 

\begin{itemize}
	\item \textbf{Iceberg.} The client executes without interruption, but the total size of the order is never visible. As such, when the order is filled by the market maker, it is immediately replenished. At some point, the client finishes executing, and the order is no longer renewed. To represent this, the replenishment probability is set to $p = 0.9$, meaning there is a 10\% chance that, after liquidity is taken, it is not renewed. Additionally, there is a small probability that the order is spontaneously cancelled by the client, given by $\nu = 0.001$ s\textsuperscript{-1}. For simplicity, we do not allow new orders after the iceberg order finishes executing, so $\mu = 0$. We let the order size be one million notional, and therefore set $\overline{\ell} = 1$.

	\item \textbf{TWAP.} The client executes at a constant pace, but orders arrive with pauses of random lengths from the perspective of the market maker. There is no instantaneous replenishment, so we set $p = 0$. The arrival process $A$ represents the renewal of the order some time after it has been consumed, and we set its intensity to $\mu = 0.05$ s\textsuperscript{-1}. As in the case of the iceberg strategy, we let $\nu = 0.001$ s\textsuperscript{-1} and $\overline{\ell} = 1$.

	\item \textbf{Full Amount.} The client places their entire order at once and never updates it, except for the possibility of cancellation. Therefore, there is no instantaneous replenishment ($p = 0$) and no arrivals ($\mu = 0$). However, the initial order size is larger than one, and we set $\overline{\ell} = 10$. As before, the cancellation rate is $\nu = 0.001$ s\textsuperscript{-1}.
\end{itemize}

\subsection{Optimal quotes} \label{sec:pricing}

\begin{figure}[t]
    \centering
    \begin{subfigure}[b]{0.5\textwidth}
        \centering
        \includegraphics[height=1.95in]{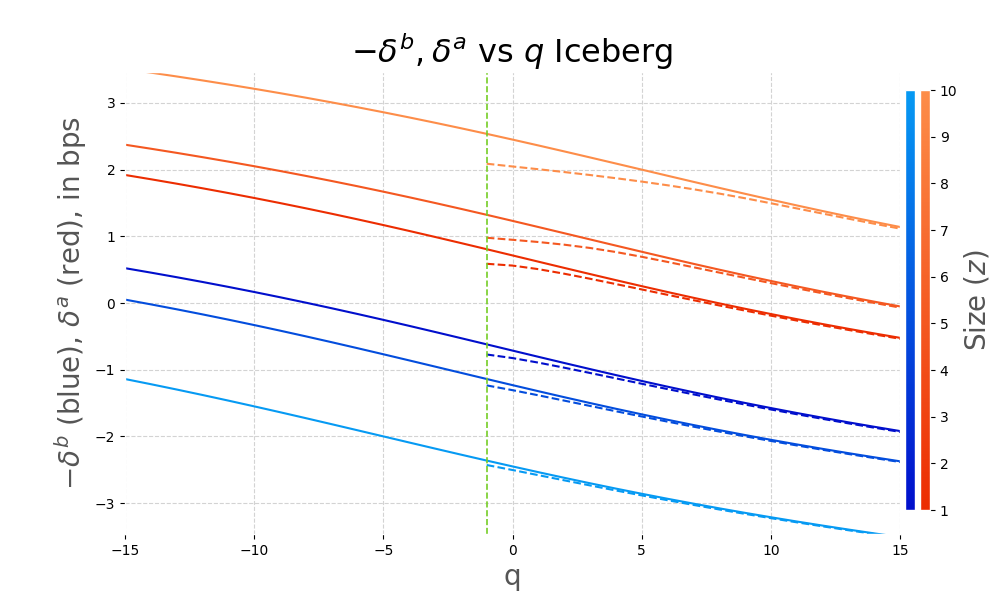}
    \end{subfigure}%
	~
    \begin{subfigure}[b]{0.5\textwidth}
        \centering
        \includegraphics[height=1.95in]{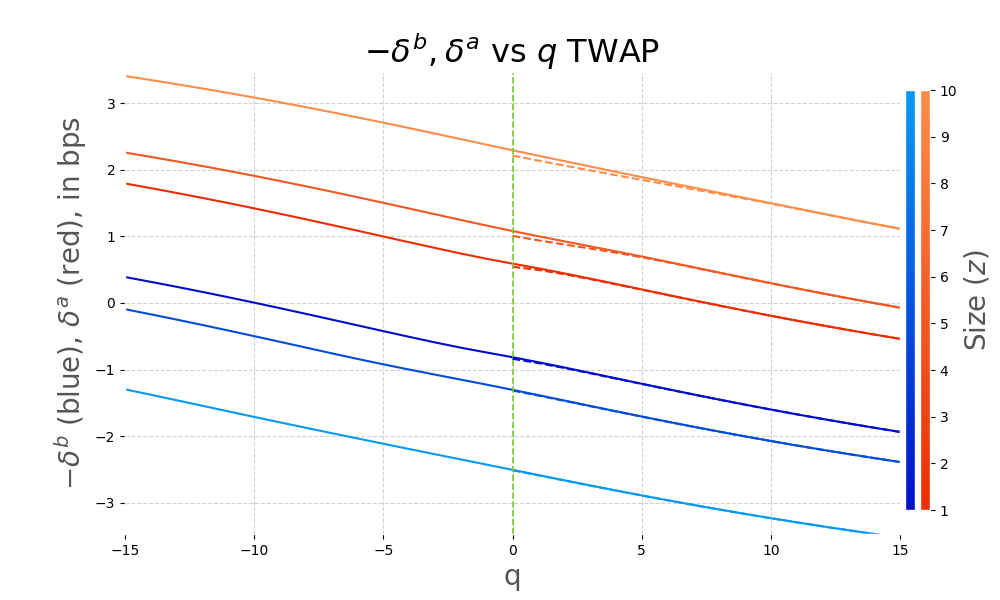}
    \end{subfigure}
    \caption{Ask and bid depths, $\delta^{b, z}(0, q, 1)$ and $\delta^{a, z}(0, q, 1)$, when an internal exchange order is present (dashed lines), and when it is not (solid lines), for $z \in \mathcal{Z}$, in the presence of an iceberg (left) and TWAP (right) client algorithm in the internal exchange.
Bid (ask) depths are shown in blue (red), with lighter shades corresponding to larger values of $z$.
Dashed lines corresponding to cases where the internal exchange order is available are shown only for values of $q$ outside the execution region, where the internal exchange order is taken.
The area to the left of the green line indicates the region where the market maker trades with the internal exchange.}
\label{fig:deltas_vs_q}
\end{figure}

In Figure \ref{fig:deltas_vs_q} we plot the optimal bid-depths $\delta^{b, z} $ and ask-depths $\delta^{a, z}$ at $t=0$ for $z\in\mathcal{Z}$ for different inventory positions of the market maker, in the case where client orders are placed at mid price ($\tilde{\rho}=0$). We chose the snapshot at $t=0$ as the strategy is approximately stationary away form the terminal time $T$.  
We observe in Figure \ref{fig:deltas_vs_q} (left panel) that, in the iceberg scenario, the ask-side quotes are lower near the execution region (i.e., when the market maker has a short position greater than one unit, $q < -1$) and the limit order is present (dashed lines), compared with the classical Avellaneda-Stoikov benchmark (solid lines). This occurs because the market maker is more willing to accumulate inventory, given the possibility of closing the position through the limit order if necessary. The effect becomes more pronounced with larger ask sizes and weaker with larger bid sizes, forming a substantial passive impact on both bid and ask prices. In contrast, this effect is marginal in the presence of a TWAP order due to the time intervals between successive renewals of the client’s orders (see right panel).

\begin{figure}[t]
    \centering
    \begin{subfigure}[b]{0.5\textwidth}
        \centering
        \includegraphics[height=2.0in]{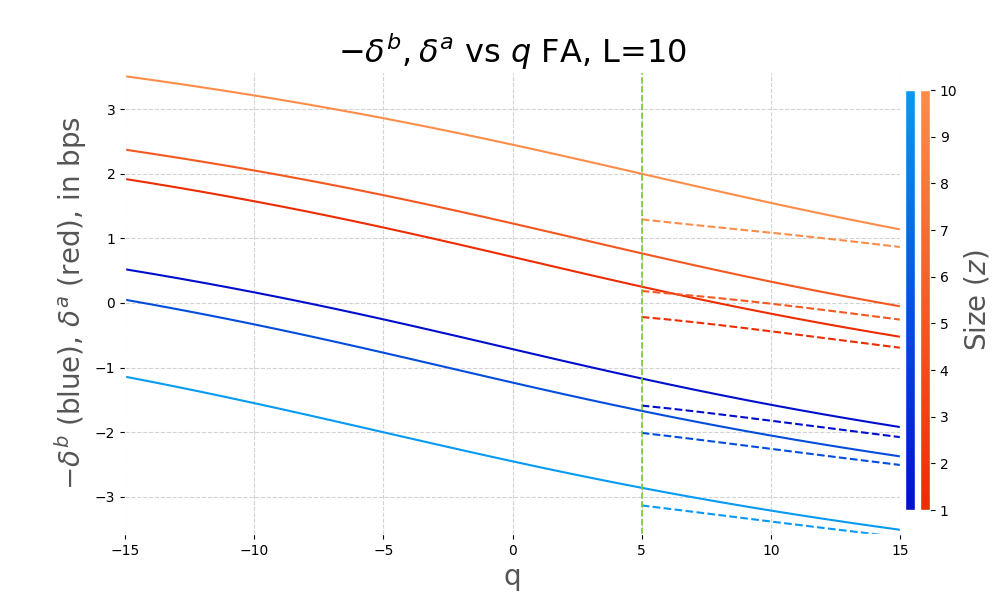}
    \end{subfigure}%
	~
    \begin{subfigure}[b]{0.5\textwidth}
        \centering
        \includegraphics[height=2.0in]{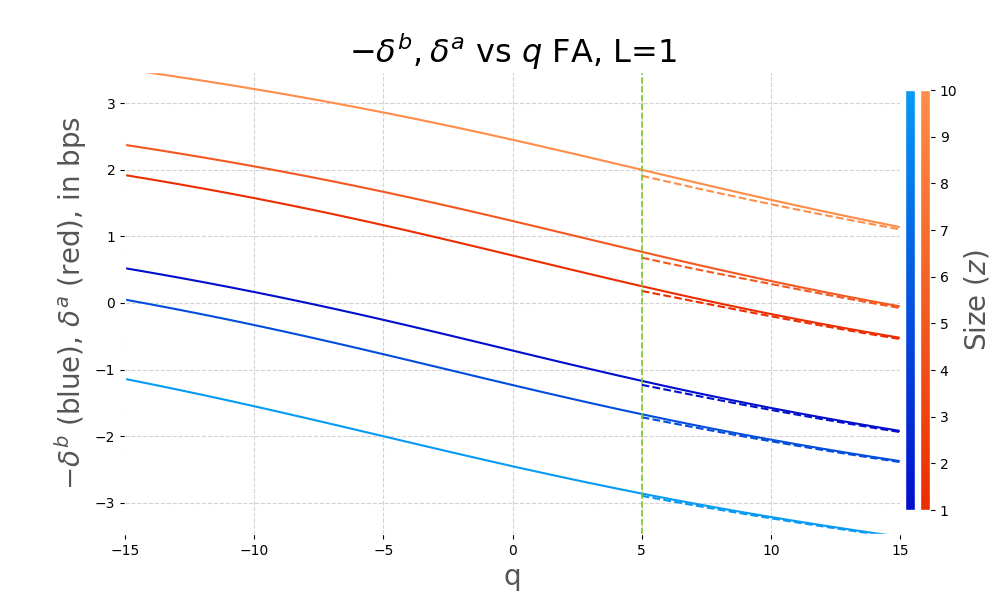}
    \end{subfigure}
    \caption{Ask and bid depths when then internal exchange order is present as in Figure \ref{fig:deltas_vs_q}, now for when the internal exchange order is of full amount type with 10 (left) and 1 (right) units remaining.} 
\label{fig:FA_deltas_vs_q}
\end{figure}

In Figure~\ref{fig:FA_deltas_vs_q}, we show the full (one-time) client order scenario for an outstanding order of 10 units (left) and one unit remaining (right). Prices adjust much more relative to the no-internal-order case when the order is full at $L = 10$, since greater available liquidity allows larger positions to be closed internally. 
Note that the trading region here is now $q<5$ due to the fact that the order will not repeat itself and the market maker can mitigate the internal execution urgency term (the fifth term on the right-hand side of  \eqref{def:value}) by taking the liquidity. This happens both when $L=10$ and $L=1$ due to the linearity of the penalty for positive $L$.

We therefore conclude that the type and magnitude of order determines the extent of price adjustment due to internal exchange orders, which can be perceived externally as passive price impact. The pricing adjustments are more pronounced when the expected remaining liquidity on the internal exchange is high.  
Since clients rarely disclose their trading strategies, a data-driven approach may be needed to infer internal order properties and decide on external price adjustments, initially assuming a standalone order and updating quotes after the first fill.

\subsection{When to trade with the internal exchange}
\begin{figure}[H] 
    \centering
    \begin{subfigure}[b]{1.0\textwidth}
        \centering
        \includegraphics[height=2.15in]{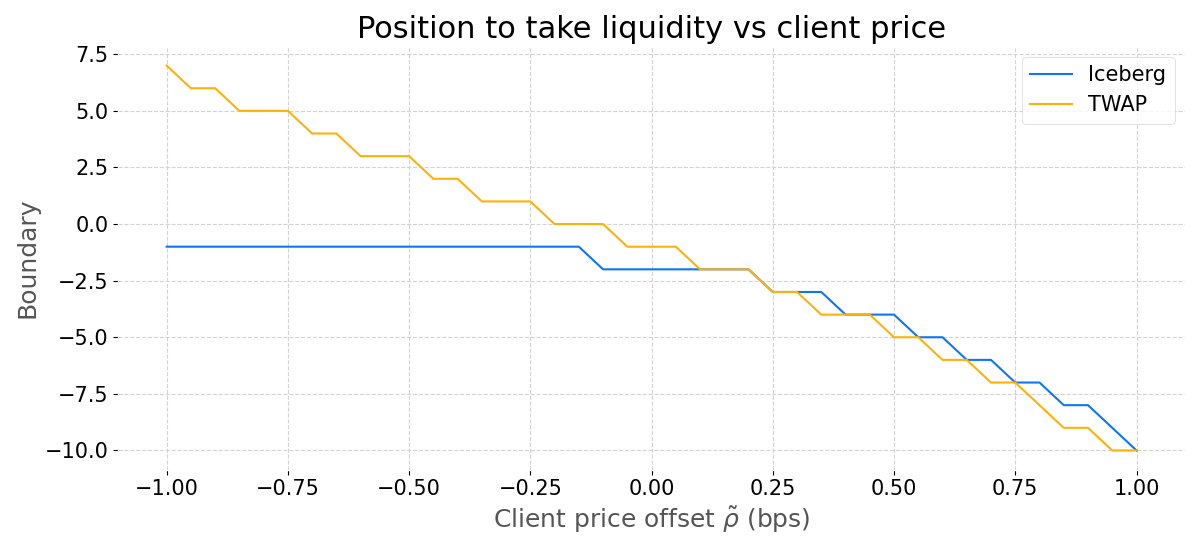}
    \end{subfigure}
    \caption{The largest position at which it is optimal to fill the internal exchange order vs. as the client's price offset $\tilde{\rho}$. The blue line illustrates the iceberg order scenario, while the orange line shows TWAP case.}
\label{fig:execution_boundaries}
\end{figure}

In Figure \ref{fig:deltas_vs_q}, we observed that, for the chosen parameters, it was optimal to trade with the internal exchange whenever the position was negative in the TWAP case, and when the position was sufficiently negative in the iceberg case. In contrast, in Figure \ref{fig:FA_deltas_vs_q}, it was optimal to trade when the position was already positive. This indicates that the optimal execution boundary depends on the model parameters. Figure \ref{fig:execution_boundaries} illustrates how this boundary changes with the client’s price $\rho$ in the iceberg and TWAP cases. Recall that, in this section, there is no fee ($\xi = 0$), so $\rho = \tilde{\rho}$.

We observe that when the client posts aggressively, with a price below the mid ($\tilde{\rho} < 0$), the market maker may be willing to trade with the internal exchange order even if doing so worsens their position (a behaviour known as risk increasing). This pattern reflects a trade-off between optimising P\&L, managing inventory costs, and accommodating the urgency of executing internal orders (as described below \eqref{def:value}). Owing to the quadratic inventory risk term, holding a small position is penalised less per unit than holding a large one; hence, the market maker may prefer to hold a small negative position rather than always remain non-negative when $\tilde{\rho}$ is large.

One of the main conclusions from Figure \ref{fig:execution_boundaries} is that clients can expect the time required to fill their orders to be more sensitive to their price offset from the mid when using a TWAP strategy than when using an iceberg order. We demonstrate this result quantitatively in Table \ref{tab:fill_times} in Section \ref{sec:comparison}. The main reason for the difference in trading regions between the iceberg and TWAP scenarios for aggressive client orders ($\tilde{\rho} < 0$) is that, at these prices, iceberg orders effectively represent a batch of existing orders that the market maker can use to make an immediate profit and to close a short position when needed. Therefore, the market maker tends to postpone consuming them until holding a sufficiently negative inventory. In contrast, TWAP orders arrive at an exponential rate and are only placed when there are no outstanding client orders. Hence, it is more profitable to consume them immediately regardless of inventory, as otherwise new orders will not arrive.   

\section{Na\"{i}ve benchmark and performance comparison}\label{sec:comparison}
In this section, we compare the performance and behaviour of the optimal strategy obtained by solving the HJBQVI \eqref{eq:HJBQVI_simplified} and using the optimal depths \eqref{eq:deltas}, with common heuristic benchmark strategies. We use the same parameters as in Section \ref{sec:strategy}. Moreover, due to the 24-hour nature of the FX market, we evaluate the optimal strategy at time $t=0$ to neglect terminal inventory constraints in our comparison. This assumption is justified, as the strategy becomes time-independent when sufficiently far from the end of the time horizon. The following algorithm describes the na\"ive benchmark strategy.

\begin{algorithm} \label{algo}
  \caption{Na\"ive benchmark strategy}\label{alg:naive}
  \begin{algorithmic}
      \State \textbf{Initiate} Set $t=0$ and $L_0=\bar\ell \geq 0$ in \eqref{def:L}. 
      \While{$t \leq T$}
    \If{the available liquidity $L_t \leq 0$}
      \State Use Avellaneda--Stoikov quotes (i.e. \eqref{def:lambda_a_b} at time zero with $L\equiv 0$),
    \Else
      \If{$Q_t < 0$}
        \State Purchase the internal exchange outstanding order and update $L_{t+\Delta} \gets L_t-1$
      \Else
        \State Use Avellaneda-Stoikov bid quotes and \textbf{adjusted} ask quotes in \eqref{def:benchmark_delta_a}.
      \EndIf
    \EndIf
    \State \textbf{Set} $t\gets t+\Delta$. 
     \EndWhile
  \end{algorithmic}
\end{algorithm}

In Algorithm \ref{alg:naive} when the market maker's position is negative and an internal exchange order is present, it is filled\footnote{In reality, this may depend on the client's posted price $\tilde{\rho}$; however, this is reasonable for aggressive and mildly passive pricing.}. If no such order exists, prices are set using the Avellaneda-Stoikov model, that is the case when there is no internal exchange ($L\equiv 0$). Prices on the ask side are adjusted according to a hubristic rule which is described below. 

We let $\delta^{a, z_{k}, AS}(0, q)$ denote the optimal half-spreads at $t = 0$ in the Avellaneda–Stoikov model, i.e., those corresponding to our model with $L \equiv 0$ and allowing for multiple order sizes (see \cite{bergault2021size}). 
In the presence of a client’s sell order of size $l$ at a price $\tilde{\rho}$ above the mid-price, the market maker inserts an order on the ask side of their pricing ladder with size $l$ at a price $\tilde{\rho} + \iota$ above mid, where $\iota > 0$ is the margin introduced by the dealer. We then compute a new volume-weighted average price (VWAP) to obtain the naïve strategy’s half-spreads.

 Let the order sizes $z_{1}<z_{2}, ...$ and let $z_{i}=\min\{z\in\mathcal{Z} : \tilde{\rho}+\iota<\delta^{a, z, AS}(0, q)\}$. The na\"{i}ve strategy's half-spreads when there is an order on the internal exchange are then given by 
$\tilde{\delta}^{a, z_{j}, BM}(0, q, l) = \delta^{a, z_{j}, AS}(0, q)$ if $\delta^{a, z_{j}, AS}(0, q) < \tilde{\rho}+\iota$, which is when $j<i$, and  for $j\geq i$, 
\begin{equation} \label{eq:benchmark_delta_a_above}
\begin{split}
	\tilde{\delta}^{a, z_{j}, BM}(0, q, l) = 
	&\frac{1}{z_{j}}\Bigg(  l \left(\tilde{\rho}+\iota\right)  + z_{i-1}\delta^{a, z_{i-1}, AS}(t, q) +\\
	&\qquad + l\sum_{r=i+1}^{j}\left(\left(\frac{z_{r-1}\delta^{a, z_{r-1}, AS}(t, q)-z_{r-2}\delta^{a, z_{r-2}, AS}(t, q)}{z_{r-1}-z_{r-2}}\right)\right) \\
	&\qquad + \sum_{r=i}^{j}\left(\left(z_{r}-z_{r-1}-l\right)\left(\frac{z_{r}\delta^{a, z_{r}, AS}(t, q)-z_{r-1}\delta^{a, z_{r-1}, AS}(t, q)}{z_{r}-z_{r-1}}\right)\right)
\Bigg)
\end{split}
\end{equation}
To summarise, the prices quoted by the market maker are, 
\begin{equation} \label{def:benchmark_delta_a}
	\delta^{a, z_{j}, BM}(0, q, l) = 
	\begin{dcases}
		\tilde{\delta}^{a, z_{j}, BM}(0, q, l) & \text{ if } l>0, \\
		\delta^{a, z_{j}, AS}(0, q)  & \text{ if } l\leq0.
	\end{dcases}
\end{equation}
The prices on the bid-side remain unchanged in the presence of the client's order, that is  $\delta^{a, z_{j}, BM}(0, q, l) = \delta^{a, z_{j}, AS}(0, q)$ (see Algorithm \ref{alg:naive}).\footnote{Alternatively, one can consider removing equivalent size from the bid ladder.}

In Table \ref{tab:pnls}, we compare the mean and standard deviation of the simulated P\&L, defined as $P\&L = X_{T} + Q_{T}S_{T}$ (see \eqref{def:value}), across different types of client orders and client pricing levels: aggressive ($\rho = -0.2$), mid ($\rho = 0$), and passive ($\rho = 0.2$). In all cases, the market maker’s initial position is zero, fees are set to $\xi = 0$, the time step is $\Delta = 0.3$ s, the time horizon is $T = 300$ seconds, and the number of simulated trajectories is $5{,}000$. We compare the performance of the optimal strategy with that of the naïve benchmark strategy using a margin of $\iota = 0.1$.
We observe that the optimal strategy consistently outperforms the naïve benchmark, and that the market maker’s profits are slightly more sensitive to the client’s pricing offset $\tilde \rho$ in the TWAP and full-amount cases than in the iceberg case. 

In Table \ref{tab:fill_times}, we observe that the expected first fill times of internal client orders under the naïve benchmark strategy are generally shorter than those under the optimal strategy. This results from the naïve strategy’s more aggressive pricing adjustments and its execution rule of always taking when the inventory is negative. In particular, the market maker readily goes short and therefore fills internal exchange orders more frequently.
We also observe that the change in the mean time to first fill under the optimal strategy is more sensitive to the client’s pricing offset $\tilde \rho$ when a TWAP order is used than when an iceberg order is used. Finally, when a full-amount order is used, the time to first fill is always zero under the optimal strategy, as the initial position of zero lies within the region where the market maker trades with the internal exchange (see Figure \ref{fig:FA_deltas_vs_q}). 

\begin{table}[H]
\centering
\begin{tabular}{cc||c|c|c}
\multicolumn{5}{c}{\textbf{P\&L (K\$)}} \\
\hline
           &   & \textbf{Aggressive: $\rho=-0.2$} & \textbf{At mid: $\rho=0$} & \textbf{Passive: $\rho=0.2$}   \\
\hline
\hline
\multirow{2}{*}{\textbf{Iceberg}}
& Optimal & 3.788 (1.62)   & 3.677 (1.6)  &  3.603 (1.6)    \\
& Na\"{i}ve & 3.384 (1.69)  &  3.36 (1.69) &  3.347 (1.69)  \\
\hline
\multirow{2}{*}{\textbf{TWAP}} 
& Optimal & 3.84 (1.65)  &  3.708 (1.61) &  3.673 (1.59)  \\
& Na\"{i}ve & 3.381 (1.5)  &  3.399 (1.53)  &  3.405 (1.56)  \\
\hline
\multirow{2}{*}{\textbf{FA}} 
& Optimal & 3.744 (2.0)  &  3.547 (2.0) &   3.354 (2.0)  \\
& Na\"{i}ve & 3.452 (1.6)  &  3.433 (1.58)  &  3.404 (1.57)  \\
\end{tabular}
\caption{Mean (standard deviation) of P\&L in thousand \$, in the presence of different client algos with passive, at mid, and aggressive placement for iceberg, TWAP and full amount (FA) scenarios.}
\label{tab:pnls}
\end{table} 

\begin{figure}[H]
    \centering
    \begin{subfigure}[b]{0.5\textwidth}
        \centering
        \includegraphics[width=\textwidth]{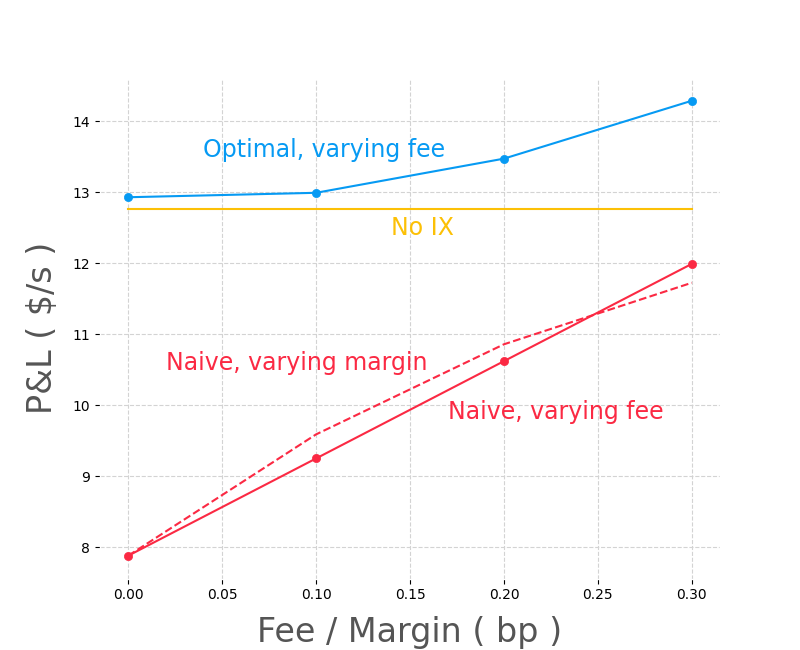}
    \end{subfigure}%
	~
    \begin{subfigure}[b]{0.5\textwidth}
        \centering
        \includegraphics[width=\textwidth]{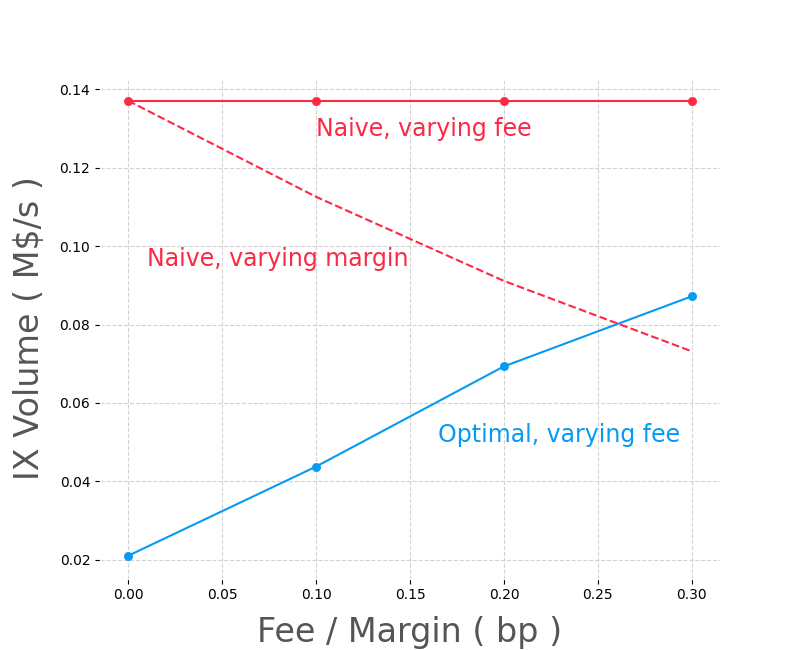}
    \end{subfigure}
    \caption{
    	P\&L (left) and internal exchange volume per second $\frac{M_{T}}{T}$ (right) for optimal market maker's strategy (blue) and na\"{i}ve benchmark strategy (red) in the presence of an iceberg as functions of fee (solid lines) or margin (dashed lines). Reference P\&L using Avellaneda-Stoikov pricing without an internal exchange is shown for comparison (orange).
    }
\label{fig:fees_and_margins}
\end{figure}

\begin{table}[H]
\centering
\begin{tabular}{cc||c|c|c}
\multicolumn{5}{c}{\textbf{Time to First Fill (seconds)}} \\
\hline
           &   & \textbf{Aggressive: $\rho=-0.2$} & \textbf{At mid: $\rho=0$} & \textbf{Passive: $\rho=0.2$}   \\
\hline
\hline
\multirow{2}{*}{\textbf{Iceberg}}
& Optimal & 21.133 (31.57)  &  47.965 (53.67) &  56.681 (63.16)  \\
& Na\"{i}ve & 5.147 (6.5)  &  7.59 (10.36) &   12.145 (18.64)  \\
\hline
\multirow{2}{*}{\textbf{TWAP}} 
& Optimal & Always zero  &  24.518 (34.69) &  68.221 (68.3)  \\
& Na\"{i}ve & 5.571 (7.7)   &  7.797 (10.24) &  12.507 (18.12)  \\
\hline
\multirow{2}{*}{\textbf{FA}} 
& Optimal & Always zero  &  Always zero &   Always zero  \\
& Na\"{i}ve & 5.309 (6.7)  &  7.402 (9.82) &   12.063 (17.34)  \\
\end{tabular}
\caption{Mean (standard deviation) of time to fill for the first internal exchange order in seconds, in the presence of different client algos with passive, at mid, and aggressive placement for iceberg, TWAP and full amount (FA) scenarios.}
\label{tab:fill_times}
\end{table}
Although execution times under the optimal strategy are slower than those of the naïve strategy, the latter substantially reduces the market maker’s P\&L. To sustain internal flow without harming performance, an additional compensation mechanism is required.
Figure \ref{fig:fees_and_margins} shows the impact of fees $\xi$ and margins $\iota$ on P\&L (left) and fill rate (right). Margins are the least effective option: while they improve the naïve strategy’s P\&L similarly to fees, they also reduce the volume of internal exchange executions. Fully compensating the naïve strategy would require a large fee, which could in practice discourage internal exchange participation, an effect not captured by the model.
In contrast, the optimal strategy benefits from fees, improving both P\&L and turnover.

\section{Conclusion} \label{sec:conclusion}
 We have introduced a model in which a market maker continuously streams prices to clients while also having the option to take liquidity from dynamic passive orders on an internal exchange. We solve for the optimal strategy using an HJBQVI and demonstrate its superior performance compared to a heuristic benchmark strategy that directly incorporates internal exchange orders into the OTC pricing ladder.

The optimal strategy defines an inventory-dependent execution threshold beyond which the market maker is willing to immediately take the internal exchange order. Otherwise, the market maker skews prices to facilitate the opposing flow. The degree of skew and the positioning of the execution threshold depend on the market maker’s risk aversion, client order depth and placement strategy, as well as on flow facilitation initiatives.

Notably, the skew mechanism revealed in our analysis highlights a potential origin of passive market impact: clients place passive orders on the internal exchange, prompting the market maker to skew prices for their OTC clients. As described in \cite{barzykin2025adverse}, these clients may then extract alpha from the resulting skew, leading the market to move.

\section*{Acknowledgments}

The views expressed are those of the authors and do not necessarily reflect the views and practices at HSBC.
The authors are grateful to Richard Anthony (HSBC) for helpful discussions and support throughout the project.

\end{document}